\documentclass{article} % For LaTeX2e
\usepackage{iclr2026_conference,times}

\usepackage{amsmath,amsfonts,bm}

\def\eqref#1{equation~\ref{#1}}
\def\1{\bm{1}}

\DeclareMathAlphabet{\mathsfit}{\encodingdefault}{\sfdefault}{m}{sl}
\SetMathAlphabet{\mathsfit}{bold}{\encodingdefault}{\sfdefault}{bx}{n}

\usepackage{url}
\usepackage{comment}
\usepackage{graphicx}
\usepackage{amsthm}
\usepackage{booktabs} % for professional tables
\usepackage{multirow}
\usepackage{todonotes}
\usepackage{tikz}
\usepackage{tikz-timing}
\usepackage{circuitikz}
\usetikzlibrary{arrows.meta}
\usetikzlibrary{shapes.geometric}
\usepackage{subfig}
\usepackage{booktabs}
\usepackage{algorithm2e}
\SetKwComment{Comment}{$\triangleright$\ }{}
\SetKwProg{Fn}{Function}{:}{}
\SetKwFunction{ConstructWitness}{ConstructWitness}
\SetKwFunction{VerifyWitness}{VerifyWitness}

\usepackage[shortlabels]{enumitem}

\usepackage{color}
\definecolor{cgreen}{rgb}{0.2,0.6,0.2}
\definecolor{darkred}{rgb}{0.4,0.0,0.0}
\definecolor{darkgreen}{rgb}{0.0,0.4,0.0}
\definecolor{darkblue}{rgb}{0.0,0.0,0.4}
\usepackage[colorlinks=true, citecolor=darkblue]{hyperref}
\hypersetup{
    linkbordercolor = {darkblue},
    urlcolor = {darkblue},
    linkcolor = {darkblue}
}

\usepackage[framemethod=TikZ]{mdframed}

\mdfdefinestyle{graybox}{%
    linecolor=white,
    outerlinewidth=0pt,
    roundcorner=3pt,
    innertopmargin=8pt,
    innerbottommargin=8pt,
    innerrightmargin=10pt,
    innerleftmargin=10pt,
    backgroundcolor=black!5!white}

\newtheorem{theoreminner}{Theorem}[section]
\newtheorem{propositioninner}[theoreminner]{Proposition}
\newtheorem{lemmainner}[theoreminner]{Lemma}
\newtheorem{corollaryinner}[theoreminner]{Corollary}
\newtheorem{observationinner}[theoreminner]{Observation}
\newtheorem{conjectureinner}[theoreminner]{Conjecture}

\newenvironment{conjecture}[1][]
  {\begin{mdframed}[style=graybox]\begin{conjectureinner}[#1]}
  {\end{conjectureinner}\end{mdframed}}

\newtheorem{definitioninner}[theoreminner]{Definition}
\newtheorem{assumptioninner}[theoreminner]{Assumption}

\newenvironment{definition}[1][]
  {\begin{mdframed}[style=graybox]\begin{definitioninner}[#1]}
  {\end{definitioninner}\end{mdframed}}

\usepackage{listings}
\usepackage{float}
\newfloat{listing}{tbp}{lop}
\floatname{listing}{Lst.}
\floatstyle{plaintop}
\restylefloat{listing}

\title{Autoformalizing Memory Device Specifications with Agents}

\author{
\textbf{Jan Ole Ernst}$^{1}$ \quad
\textbf{Dmitri Michelangelo Saberi}$^{1}$ \quad
\textbf{Derek Christ}$^{2,3}$ \\
\textbf{Thomas Zimmermann}$^{2}$ \quad
\textbf{Rajath Salegame}$^{1}$ \quad
\textbf{Suhaas M. Bhat}$^{1}$ \\
\textbf{Stanislav Levental}$^{1}$ \quad
\textbf{Thomas Dybdahl Ahle}$^{1}$ \quad
\textbf{Matthias Jung}$^{2,3}$ \\
\\
$^{1}$Normal Computing \\
$^{2}$Fraunhofer Institute for Experimental Software Engineering (IESE), Kaiserslautern, Germany \\
$^{3}$University of Würzburg, Germany \\
\\
\texttt{\{jan,dmitri\}@normalcomputing.com} \\
\texttt{thomas.zimmermann@iese.fraunhofer.de} \\
\texttt{\{derek.christ,m.jung\}@uni-wuerzburg.de}
}
\newcommand{\timemeasure}[4]
{       \draw [red,semithick] ($ (#1) - (-0.1,0) $) -- ($ (#1) - (-0.1,#3) -(0,1) $);       \draw [red,semithick] ($ (#2) - (-0.1,0) $) -- ($ (#2) - (-0.1,#3) -(0,1) $);       \draw [red,semithick,<->] ($ (#1) - (-0.1,#3) $) -- ($ (#2) - (-0.1,#3) $) node [below,midway] {#4};
}

\iclrfinalcopy % Uncomment for camera-ready version, but NOT for submission.
\begin{document}

\maketitle

\begin{abstract}
The primary goal of Design Verification (DV) is to ensure that a proposed chip design implementation (either in code, or physical form) exactly matches its specification and is free of functional errors in order to avoid costly re-designs. Achieving this often demands extensive manual interpretation, translating the specification document into a formal, testable representation. While AI has made progress in DV, current approaches typically focus on narrow, isolated tasks rather than full end-to-end specification compliance of modern chip designs, failing to capture the complexity of real-world verification. 
Our method automatically formalizes natural language memory chip specifications, for industry relevant Dynamic Random Access Memory (DRAM) standards, into a formal representation called \textit{DRAMPyML} that can be used for downstream DV tasks like the generation of SystemVerilog assertions, stimulus, and functional coverage. We also release our benchmarking dataset, \textit{DRAMBench}, which can be used to evaluate the evolution of model capabilities (and new approaches) at hardware autoformalization.
\end{abstract}

%\section*{Acknowledgements}

%\textbf{Do not} include acknowledgements in the initial version of
%the paper submitted for blind review.

%If a paper is accepted, the final camera-ready version can (and
%usually should) include acknowledgements.  Such acknowledgements
%should be placed at the end of the section, in an unnumbered section
%that does not count towards the paper page limit. Typically, this will 
%include thanks to reviewers who gave useful comments, to colleagues 
%who contributed to the ideas, and to funding agencies and corporate 
%sponsors that provided financial support.

\section{Introduction}
Designing complex semiconductor devices such as memory chips, processors, and system-on-chip (SoC) components is a knowledge-intensive process where correctness is paramount. A significant bottleneck in modern hardware design is the verification and validation phase, where a comprehensive \textit{testbench} is built to ensure that a register-transfer level (RTL) model adheres to its design specification. Industry-wide surveys estimate that verification consumes over half of the chip design lifecycle on average \citep{functional-verification-survey}. This enormous verification burden arises from the need to manually translate design intents (often described in natural language documents) into checkable properties, models, and other testbench components. The lack of automated tools and workflows to streamline this translation process contributes to extended design cycles and potential spec misinterpretation, ultimately risking design flaws slipping into silicon hardware.

The importance of bridging the gap between informal natural language specs and formal design representations has grown in recent years. Hardware designs have been evolving quicker than ever in functionality and complexity; growing device density is leading to new kinds of physical bugs (e.g., the row hammer attack for memory devices \citep{rowhammer}) and AI's growing power needs have given rise to novel unconventional architectures \citep{golden-age-architectures}. On the other hand, the specifications that describe these designs (such as JEDEC standards - the Joint Electron Device Engineering Council: provides interoperability standards common for memory chips) are written in natural language, often supplemented by timing diagrams and register map tables. Building complex verification collateral based only on a natural language specification is challenging. Ambiguities in describing device behavior in unusual command sequences or states can cause subtle bugs that are hard to detect using the natural language specification as ground truth.

An \textit{autoformalization system for hardware}---one that automatically converts a natural language spec into an unambiguous formal model–-could dramatically streamline verification and further ``shift left'' the design/verification loop \citep{shift-left}. Though not explicitly described as autoformalization, many methods exist to turn natural language specs into (formal) SystemVerilog Assertions (SVAs) using Large Language Models (LLMs) \citep{assertllm, assertionforge, assertcoder}. SystemVerilog Assertions (SVAs) specify expected device properties (e.g., timing constraints) that can be checked in simulation or formally proven using verification tools such as SAT/SMT solvers \citep{assertions-sv}. Formal proving can however be a huge bottleneck for complex designs with SAT solvers running for hours to establish whether an assertion is provably satisfied making it difficult to embed this in AI workflows. While SVAs are widely used in industrial verification, directly generating them from natural language specifications remains difficult: even state-of-the-art models frequently misinterpret specific protocol details, producing incorrect or irrelevant assertions \citep{shih2025flagformalllmassistedsva, assertionforge}. Furthermore, SVAs alone are insufficient for building complete verification environments, as they do not directly support other testbench collateral such as stimulus generation or functional coverage, which actively drive the design under test through various scenarios to quantitatively measure which edge cases, and protocol features have been tested.

We therefore propose to formalize memory specifications into a more general, abstract representation using the \textit{DRAMPyML} language introduced in \cite{drampyml} with an iterative autoformalization \textit{agent}. The resulting formal model can be translated downstream into several different testbench components. We consider the following DRAM standards; DDR2/3/4/5, LPDDR2/3/4/5, GDDR5/6/7 and HBM2/3. These cover the major DRAM categories, including general-purpose systems, low-power system, graphics and AI Accelerators.
% DDR for general-purpose systems, LPDDR for mobile and low-power devices, GDDR for graphics and accelerators, and HBM for high-bandwidth workloads such as GPUs and AI accelerators. 

In summary, our contributions are as follows:\vspace{-0.2cm}
\begin{enumerate}[leftmargin=20pt, labelsep=.6em]
  \setlength\itemsep{0em}
    \item We examine theoretical properties of the DRAMPyML models, and justify empirically an efficient means of verifying (i) symmetry properties and (ii) correctness of generated formalizations  (Section~\ref{sec:background}).
    \item We present an agentic framework (Section~\ref{sec:method}), and show that it achieves higher scores than baseline methods, including perfect structural descriptions for some specifications and analyze the token (cost) vs. performance tradeoff for various models (Section~\ref{sec:results}).
    \item We release (https://github.com/normal-computing/DRAMBench) an open source benchmark for hardware autoformalization for 12 memory standards across DDR, LPDDR, HBM, and GDDR device families (Section~\ref{sec:dataset}).

\end{enumerate}

The paper is structured as follows: Section~\ref{sec:related-work} will present the related work. In Section 3 the required Background on DRAMpyML and Petri Nets is discussed. Our Auto-formalization approach is presented in Section 4, while Section 5 presents the results of this method.

%There are several benefits to SVAs as the choice of formal representation, namely (i) it is straightforward to build a dataset of (natural language statement, formal property) tuples \cite{fveval}, (ii) SVAs are used with formal techniques in hardware verification already  (e.g. Z3, SAT solvers), and (iii) there is a great body of literature in using LLMs for SVA generation \cite{assertcoder, assertllm, assertionforge}. 

\begin{figure}
    \center
    \subfloat[DRAM timing diagram showing the timing constraint $t_{RCD}$ between an \texttt{ACT} and a \texttt{RD} command.]{%
        \input{command_sequence}
        \label{fig:command_sequence}
    }
    \hspace{10pt}
    \subfloat[Petri net showing timing dependencies after firing \texttt{ACT} of Bank 0.]{%
        \includegraphics[width=0.28\linewidth]{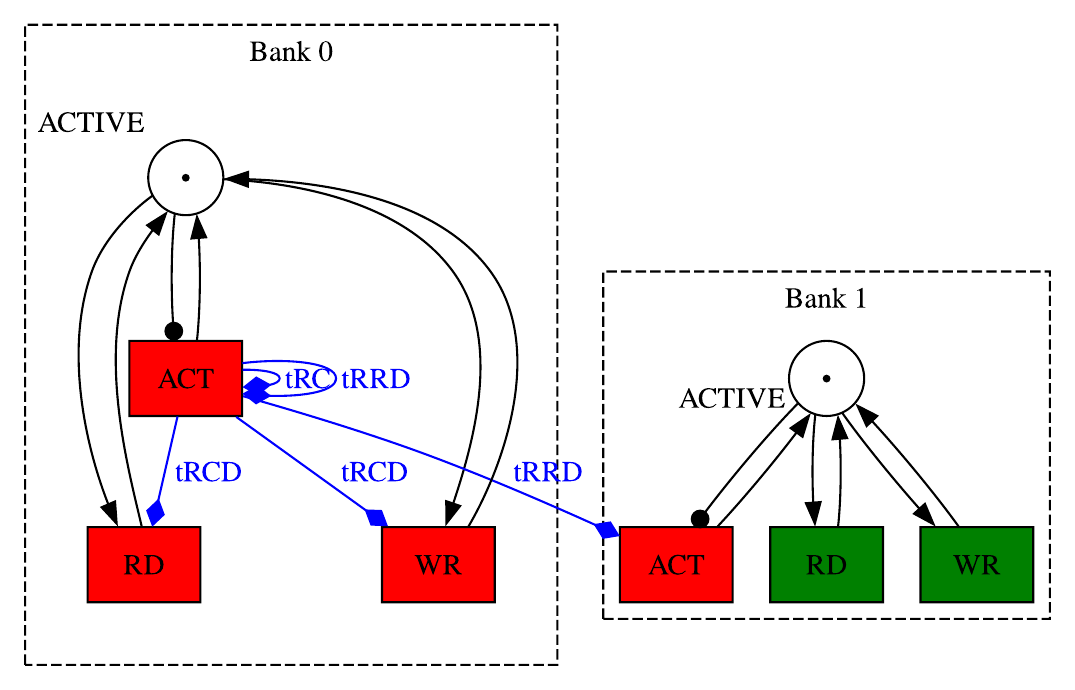}
        \label{fig:petri_net_timing}
    }
    \caption{DRAM timing constraint (a) $t_{RCD}$ between an \texttt{ACT} and a \texttt{RD} to bank 0 with Petrinet (b).}
    \vspace{-0.5cm}
\end{figure}

\section{Related Work}\label{sec:related-work}

Several recent works use LLMs to translate informal hardware specifications into formal properties, most commonly SystemVerilog Assertions (SVAs) \citep{assertllm, assertionforge, assertcoder}. While this supports automated property checking, assertion-centric approaches do not yield complete behavioral models or a structured intermediate representation, and often generate large numbers of syntactically invalid or specification-misaligned assertions due to ambiguity in natural language. Recent work highlights these limitations: FLAG \citep{shih2025flagformalllmassistedsva} shows that direct LLM-based SVA generation struggles with unstructured communication protocol specifications, motivating hybrid frameworks that rely on formal templates and diagram-based filtering.

Autoformalization has also been studied in mathematics, where LLMs convert informal statements into machine-checkable code in theorem provers such as Lean4 \citep{gulatietal2024lean4bench}. These benchmarks demonstrate both the promise and difficulty of formalization, but focus on mathematical logic rather than the concurrent, timing-rich operational models required in hardware.

Importantly, none of the existing approaches have tackled the autoformalization of full memory chip standards such as DDR, which involve long specifications, complex command interactions, and intricate timing constraints. Indeed, the lack of open source professional-grade memory Verification IP (VIP) has led to the notable omission of such standards from so-called \textit{AI for SVA generation} benchmarks (see \citep{assertcoder, assertionforge, assertllm}). In contrast, our work targets full-specification autoformalization of modern memory standards into structured executable models, enabling downstream generation of diverse verification collateral beyond isolated assertions. There are various upsides in using DRAMPyML to summarize the formal requirements described in a specification (as opposed to, say, formalizing directly into SVAs). The algorithm to generate SVAs from DRAMml, as introduced in \citet{dramml-svas}, extends directly to DRAMPyML. Moreover timed command traces (defined in detail in \ref{sec:drampyml-math}) correspond directly to transaction-level \textit{SystemVerilog stimulus} (see, e.g. \citet{spear_systemverilog}). 
    \vspace{-0.2cm}
\section{Background}\label{sec:background}
    \vspace{-0.2cm}
%\subsection{DRAM Devices}

% 1T1C here
In what follows we give an overview of DRAM devices and explain our custom markup language for modeling such devices. A DRAM device is a random-access memory (RAM) that uses a pair of transistors and a capacitor for each memory cell.
The JEDEC defines a large set of DRAM standards, roughly one per year (see App. Fig.~\ref{fig:jedec_standards}). 
%\begin{figure}
%    \centering
%    \includegraphics[width=0.9\linewidth]{iclr2026/figures/jedec_standards.pdf}
%    \caption{JEDEC standard releases over time(~\cite{junkra_19}).}
%    \label{fig:jedec_standards}
%\end{figure}
DRAM devices are controlled by command sequences per their JEDEC standard.
Before data can be accessed, a row in a bank must be opened with an \texttt{ACT} command. Subsequent \texttt{RD} and \texttt{WR} commands operate on this active row, issuing bursts over the data bus. To access a different row, the bank must first be closed and the bitlines restored using a \texttt{PRE} command. Correct operation requires adherence to numerous timing constraints specified by the standard; for instance, after issuing an \texttt{ACT} command, the memory controller must wait at least $t_{RCD}$ until it can issue a \texttt{RD} command for the same row, as shown in Fig.~\ref{fig:command_sequence}.
%These cells are connected to their local wordline (LWL) and local bitline (LBL) and are organized as rows and columns into memory arrays that, in turn, make up a complete memory bank.
%In each bank, only one row can be actively fetched into the so-called primary sense amplifiers (PSAs) from that the secondary sense amplifiers (SSAs) burst a complete memory access over the data bus.
%The general architecture of a DRAM device is shown in App. Figure~\ref{fig:dram}.
%DRAM devices are controlled using DRAM commands: before data can be accessed, a row in the DRAM bank needs to be activated using an \texttt{ACT} command.
%After that, \texttt{RD} and \texttt{WR} commands can read and write data into the currently opened DRAM row.
%To access another row, the bitlines of the memory array need to be precharged using the \texttt{PRE} command.
%The DRAM commands are defined by the specific JEDEC standard for the respective memory type.
%Most DRAM commands must adhere to the timing constraints specified in the JEDEC standard:
%For example, after issuing an \texttt{ACT} command, the memory controller must wait at least $t_{RCD}$ until it can issue a \texttt{RD} command for the same row, as shown in Figure~\ref{fig:command_sequence}.

% Modeling using Petri nets
%With recent standards such as LPDDR6 and DDR5, the DRAM protocol became increasingly complex, making it difficult to understand just from the specification.
\citet{junkra_17,junkra_19} introduced DRAMml, a timed Petri net–based modeling technique for DRAM protocols, capable of representing increasingly complex standards such as DDR5. Fig.~\ref{fig:petri_net_timing} shows the Petri net of two DRAM banks after an \texttt{ACT} in \textit{Bank 0}, where blue timing arcs indicate active timing constraints that block other commands. Recent work extends DRAMml to a Python-based model, DRAMPyML~\citep{drampyml}.
% , which is introduced in the following subsection.

\subsection{DRAMPyML}\label{sec:drampyml}

DRAMPyML is a modeling approach for DRAM standards using timed Petri nets ~\citep{drampyml} addressing the challenges emerging from formalizing increasingly complex DRAM protocols. Like DRAMml~\citep{dramml}, it uses timed Petri nets to capture structure and constraints. Unlike DRAMml, DRAMPyML leverages Python for greater flexibility and direct executability while still supporting source code generation and verification. Listing~\ref{lst:drampyml} shows how such a Petri net is iteratively built up using DRAMPyML.
\begin{listing}
    \caption{Definition of the general Petri net graph structure.}
    \vspace{-0.4cm}
    \begin{lstlisting}[language=Python]
g = rx.PyDiGraph()
for rank in range(numberOfRanks):
    for bank in range(numberOfBanks):
        # ACTIVE place with an associated bank coordinate:
        p_active = g.add_node(Place(ACTIVE, bank_coord))
        # PREA transition with an associated rank coordinate:
        t_prea = g.add_node(Transition(PREA, rank_coord))
        # Reset arc from ACTIVE to PREA:
        g.add_edge(p_active, t_prea, ResetArc())
\end{lstlisting}
\label{lst:drampyml}
    \vspace{-0.4cm}
\end{listing}
In each level of hierarchy, the respective places, transitions and arcs are added to the graph. The timing dependencies are similarly defined at this level.
However, due to the large number of possible command transitions, the timing arcs are generated in a second step, as shown in Listing~\ref{lst:pyml_timing_constraint}.
\begin{listing}
\caption{Definition of the $t_{RCD}$ dependency for a bank.}
    \vspace{-0.4cm}
\begin{lstlisting}[language=Python]
CommandTimingConstraint(
    intra_bank, [ACT], [RD, WR, RDA, WRA], tRCD
)
\end{lstlisting}
    \vspace{-0.4cm}
\label{lst:pyml_timing_constraint}
\end{listing}
This example applies the timing constraint $t_{RCD}$ between the \texttt{ACT} command and the column commands \texttt{RD}, \texttt{RDA}, \texttt{WR} and \texttt{WRA} for this specific memory bank.
This is because the listed commands are interpreted as a Cartesian product of the two lists 
% $\{ACT\} \times \{RD,RDA,WR,WRA\}$ 
and generates a timing arc for every combination of commands.
% With a general overview over the modeling of DRAMs using timed Petri nets, the following subsection defines the mathematical definitions of the introduced concepts.

\subsubsection{Formal definitions of DRAMPyML Constructs}\label{sec:drampyml-math}
With a general overview over the modeling of DRAMs using timed Petri nets, we now introduce the mathematical definitions of the introduced concepts, starting with the DRAM Petri net.

\begin{definition}[DRAM Petri net]\label{def:dram-petri-net}
 A DRAM Petri net is a tuple $\mathcal{N} = (P, T, A, I, S, \mathcal{T}, W, M_0)$:
 \begin{enumerate}[leftmargin=20pt, labelsep=.6em]
  \setlength\itemsep{0em}
     \item $P = P_{B} \cup P_R$ is a finite set of places, partitioned into:
     \begin{enumerate}
         \item Bank places $P_B : b \in \mathcal{B}$ for bank set $\mathcal{B}$
         \item Rank (channel, pseudochannel) places $P_R : r \in \mathcal{R}$ for rank (channel, pseudochannel) set $\mathcal{R}$
    \end{enumerate}
         \item $T$ is a finite set of transitions (commands), where each $t \in T$ has an associated coordinate $\text{coord}(t) \in \mathcal{B} \cup \mathcal{R}$
         \item $A \subseteq (P \times T) \cup (T \times P)$ is the set of regular arcs (determines the allowed state flow)
         \item $I \subseteq P \times T$ is the set of inhibitor arcs (prohibits certain transitions from places)
         \item $S \subseteq P \times T$ is the set of reset arcs (clears tokens from source place) 
         \item $\mathcal{T} \subset T \times T \times \mathbb{N}$ is the set of timed arcs (encoding required delays between commands; untimed nets don't include this set)
         \item $W : A \cup I \to \mathbb{N}$ is the arc \textit{weighting} map 
         \item $M_0 : P \to \mathbb{N}$ is the initial marking 

 \end{enumerate}
\end{definition}

The key differences between Definition \ref{def:dram-petri-net} and a regular (timed-inhibitor-reset) Petri net are the coordinate decompositions attached to the places and transitions. Note that the memory devices under consideration (HBM, DDR, GDDR, notably excluding 3DS devices) have different names for the top of their hierarchy, i.e. pseudochannel, channel, rank, and logical rank.
% , which function similarly. 
They all serve the same purpose, so without loss of generality, we will use \textit{rank}. A DRAM Petri net is parameterized by number of banks $|\mathcal{B}|$ and number of ranks $|\mathcal{R}|$; we will make this parameterization explicit when necessary. Note further that some memory devices (e.g. DDR5) have a \textit{bank group} $\mathcal{B}_g$ construct to admit further parallelism, in which case the bank set $\mathcal{B}$ is further decomposed as $\mathcal{B} = \mathcal{B}_b \times \mathcal{B}_g$.

\textbf{Command traces} Transitions are commands that change the state (or \textit{marking}) of the Petri net. The different kinds of arcs in the Petri net, with their types and weights, fully determine which transitions are allowed (with respect to a marking $M$ and a time $\tau$), and if they are, how the marking will change (i.e. how the tokens will move). We call the map that checks whether a transition is enabled
$\operatorname{en} : \mathbb{N}^P \times T \times \mathbb{N} \to \{0,1\}.$
That is, $\operatorname{en}(M, t, \tau) = 1$ iff the transition $t$ can \textit{fire} with respect to marking $M$ at time $\tau$. Interpreting these as legal DRAM command sequences, we define:
\begin{definition}[Command traces]\label{def:command-traces} 
$\operatorname{Tr}_k(\mathcal{N}) \subset T^k$ denotes the set of allowed command traces of length $k$, ignoring timing. That is, for $(t_1, \dots, t_k) \in \operatorname{Tr}_k(\mathcal{N})$, there exist times $0 = \tau_1 < \cdots < \tau_k$ such that $((t_1, \tau_1), \dots, (t_k, \tau_k))$ can fire. By choosing the \textbf{minimum} firing time $\tau^{\operatorname{min}}_i$ for each transition $t_i$, we end up with the timed command traces $$\operatorname{Tr}^{\tau}_k(\mathcal{N})= \{((t_1, \tau^{\operatorname{min}}_1), \dots, (t_k, \tau^{\operatorname{min}}_k)) : \text{ each }t_i \text{ can fire }\}\subset T^k \times \mathbb{N}^k$$
\end{definition}

DRAM Petri net representations are not unique; they can encode all the behavior mandated by the specification in different ways (e.g. by changing arc weighting and including different inhibitors). However, command traces are unique: a sequence is either permitted or disallowed in the spec. Hence what matters are \textit{trace-equivalence classes} of DRAM Petri nets, which we define as follows:

\begin{definition}[Trace equivalence]\label{def:trace-equivalence}
DRAM Petri nets $\mathcal{N}_1$ and $\mathcal{N}_2$ are (timed) trace equivalent if $\operatorname{Tr}_k^{(\tau)}(\mathcal{N}_1) = \operatorname{Tr}_k^{(\tau)}(\mathcal{N}_2)$ for all $k$. They are $L$-bounded (timed) trace equivalent if the prior condition holds for $k \leq L$.
\end{definition}

\textbf{Finding Trace Equivalence}
Proving trace equivalence is interesting both from the standpoint of verifying AI outputs and from the perspective of using DRAMPyML models in a real-world DV flow. However, since $|T|$ scales as $O(|\mathcal{B}||\mathcal{R}|)$, the standard algorithm for finding $\operatorname{Tr}_k(\mathcal{N})$ (cf. \cite{drampyml}) runs in $O((|\mathcal{B}||\mathcal{R}|)^k)$ time. Considering the size of modern devices (e.g. for DDR5, $|\mathcal{B}| = 32$), trace equivalence checking is practically limited due to the state space explosion, caused by bank- and rank-parallelism. However, under symmetries and technical properties that end up holding for DRAM, short sequences are sufficient to prove equivalence:

\begin{conjecture}[Minimal configuration equivalence]\label{thm:min-config}
Let $\mathcal{N}_1(B,R)$ and $\mathcal N_2(B,R)$ be two untimed DRAM Petri nets satisfying strong conditions held by ground truth that are efficiently verifiable (e.g. having bank/rank symmetry, arc weights linear in $B$, short setup traces to different configurations due to modes being disjoint, etc.). Then, they are trace equivalent for any $B,R$ iff the following equalities hold:
\[
\operatorname{Tr}_{4}(\mathcal N_1(1,1))=\operatorname{Tr}_{4}(\mathcal N_2(1,1))
\quad\text{and}\quad
\operatorname{Tr}_{4}(\mathcal{N}_1(2,1))=\operatorname{Tr}_{4}(\mathcal N_2(2,1)).
\]
That is, it suffices to check length $4$ traces for $1$ and $2$ bank configurations to prove full equivalence.
% Moreover, 
\end{conjecture}\vspace{-0.3cm}
While we do not present a complete proof of Conjecture \ref{thm:min-config}, we validate it empirically by bounded depth counterexample search. Concretely, we attempt to find a trace $\sigma\in T^k$ for $k\le 4$ that is enabled in exactly one of $\mathcal N_{\mathrm{gr}}(2,1)$ and $\mathcal N_{\mathrm{mut}}(2,1)$. To this end, we conducted exhaustive mutation testing on more than $1000$ injected bugs in ground-truth Petri nets, comparing $\mathcal{N}_{\mathrm{gt}}$ with mutated versions $\mathcal{N}_{\mathrm{mut}}$. The mutations included small semantics-altering edits—such as removing an inhibitor arc, perturbing an arc weight, or modifying the coordinate predicate of an arc—as well as combinations of these changes. Each mutation reflects realistic faulty-generation scenarios while preserving easily verifiable structural properties, such as bank symmetry.  We found that Conjecture \ref{thm:min-config} held in all cases: traces of length $\leq 4$ witnessed \textit{all} introduced bugs. We hope to include a full proof in forthcoming work.

Notably we have avoided evaluating the timing behavior of a DRAM Petri net, which we evaluate separately using Equation \ref{eq:timing-recall} as will be explained in Sec \ref{sec:evaluation}.

% Add listing
\vspace{-0.2cm}
\section{Specification to Petri net Autoformalization Method}\label{sec:method}
\vspace{-0.2cm}
    % \subsection{Spec$\to$ Petri net Autoformalization}
     \makeatletter
\pgfdeclareshape{document}{
  \inheritsavedanchors[from=rectangle] % this is nearly a rectangle
  \inheritanchorborder[from=rectangle]
  \inheritanchor[from=rectangle]{center}
  \inheritanchor[from=rectangle]{north}
  \inheritanchor[from=rectangle]{south}
  \inheritanchor[from=rectangle]{west}
  \inheritanchor[from=rectangle]{east}
  % ... and possibly more
  \backgroundpath{% this is new
    % store lower right in xa/ya and upper right in xb/yb
    \southwest \pgf@xa=\pgf@x \pgf@ya=\pgf@y
    \northeast \pgf@xb=\pgf@x \pgf@yb=\pgf@y
    % compute corner of ``flipped page''
    \pgf@xc=\pgf@xb \advance\pgf@xc by-5pt % this should be a parameter
    \pgf@yc=\pgf@yb \advance\pgf@yc by-5pt
    % construct main path
    \pgfpathmoveto{\pgfpoint{\pgf@xa}{\pgf@ya}}
    \pgfpathlineto{\pgfpoint{\pgf@xa}{\pgf@yb}}
    \pgfpathlineto{\pgfpoint{\pgf@xc}{\pgf@yb}}
    \pgfpathlineto{\pgfpoint{\pgf@xb}{\pgf@yc}}
    \pgfpathlineto{\pgfpoint{\pgf@xb}{\pgf@ya}}
    \pgfpathclose
    % add little corner
    \pgfpathmoveto{\pgfpoint{\pgf@xc}{\pgf@yb}}
    \pgfpathlineto{\pgfpoint{\pgf@xc}{\pgf@yc}}
    \pgfpathlineto{\pgfpoint{\pgf@xb}{\pgf@yc}}
    \pgfpathlineto{\pgfpoint{\pgf@xc}{\pgf@yc}}
 }
}
\makeatother

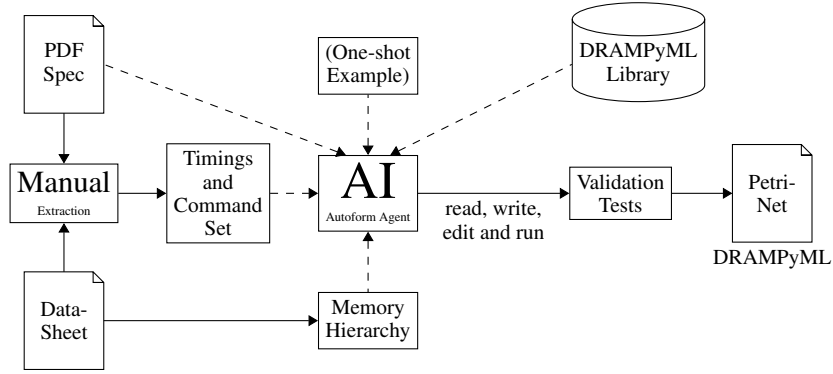
\begin{figure}[!t]
    \vspace{-0.2cm}
    \centering
    \begin{tikzpicture}[scale=0.8, transform shape]
        \draw node[draw, align=center](ai){\Huge AI\\\tiny Autoform Agent};
        \draw[dashed, Triangle-](ai.north) -- ++(0,1) coordinate(ose);
        \draw(ose) node[draw, align=center, anchor=south](){(One-shot\\Example)};
        \draw[dashed, Triangle-](ai.60) -- ++(3,1.5) coordinate(lib);
        \draw(lib) node[draw, shape border rotate=90, cylinder, anchor=west, aspect=0.25, align=center](){DRAMPyML\\Library};
        \draw[dashed, Triangle-](ai.120) -- ++(-4,1.5) coordinate(spec);
        \draw(spec) node[document, draw, anchor=east, align=center, minimum width=1.3cm,  minimum height=1.6cm](spec){PDF\\Spec};
        \draw[dashed, Triangle-](ai.270) -- ++(0,-1) coordinate(hir);
        \draw(hir) node[align=center, anchor=north, draw](mh){Memory\\Hierarchy};
        \draw[Triangle-](mh.west) -- (mh.west-|spec.east) node[document, draw, anchor=east, align=center, minimum width=1.3cm, minimum height=1.6cm](ds){Data-\\Sheet};
        \draw[-Triangle](ai.east) -- ++(2.5,0) node[midway, align=center, anchor=north](read){read, write,\\edit and run} node[draw, align=center, anchor=west](val){Validation\\Tests};
        \draw[-Triangle](val.east) -- ++(1,0) node[document, draw, anchor=west, align=center, minimum width=1.3cm,  minimum height=1.6cm](pn){Petri-\\Net};
        \draw(pn.south) node[anchor=north](){DRAMPyML};
        \path(spec.south|-ai.west)  node[align=center, draw](llm){\Large Manual\\\tiny Extraction} (ds.north);
        \path(llm.east) -- node[draw, midway, align=center](tcs){Timings\\and\\Command\\Set} (ai.west);
        \draw[-Triangle](ds.north) -- (llm.south);
        \draw[-Triangle](spec.south) -- (llm.north);
        \draw[-Triangle](llm.east) -- (tcs.west);
        \draw[-Triangle, dashed](tcs.east) -- (ai.west);
    \end{tikzpicture}

    \caption{Petri net formalization pipeline illustrating the artifacts available to the agent. The autoformalization agent has access to the DRAMPyML codebase, PDF specification, optionally to example Petri nets, timing/command definitions, and a memory hierarchy. It iteratively uses MCP tools and LLM APIs to read artifacts, generate or modify code, and run validation until all checks are satisfied. }
    \label{fig:autoform_pipeline}
    \vspace{-0.5cm}

\end{figure}
We now introduce the autoformalization pipeline, as illustrated in Fig.~\ref{fig:autoform_pipeline}, which depicts how chip specification artifacts are progressively transformed into an executable and validated Petri net model. The process begins with the primary documentation sources, namely a JEDEC PDF specification and the corresponding vendor datasheet. These artifacts are the authoritative description of the protocol and provide the details required for formal modeling. While it is possible to extract the command set, timing parameters (which may merely given as a range in JEDEC) and memory hierarchy from the PDF specification using LLMs, it does not represent a significant speedup of the workflow and will as such be addressed in future work. Here we are looking at the ability of the autoformalization agent to generate a Petri net which closely resembles the specification.

\vspace{-0.3cm}
\subsection{Autoformalization Agent} 
\vspace{-0.3cm}
Once the timing/command set and memory hierarchy have been obtained, they are passed to an autoformalization agent embedded in an interactive environment. This agent has access to the extracted protocol information, the original PDF specification, the DRAMPyML source code, and optionally example Petri net models. The agent iteratively consults these resources, calling MCP servers and an LLM API to read specification sections, write or edit Petri net code, and execute validation tests. This loop continues until the generated Petri net satisfies the agents structural consistency test and can generate command sequences and timing correctness.
% , at which point the model is considered complete.

\textbf{Prompting strategies} We additionally distinguish between two prompting strategies for autoformalization: an agentic iterative approach and a one-shot variant. While both methods operate on the same set of inputs, the agent benefits from repeated interaction with the environment, allowing the model to incrementally refine the Petri net through tool use (glob, grep, bash, write) and validation feedback. Notably the agent has access to the DRAMPyML library, hence with all of its associated algorithms (e.g. command trace generation/testing) to probe its Petri net and we prompt it to run validation tests on the graph to ensure it is well formed (no orphaned nodes) and has no deadlock states as well as generating command sequences to cross-check with the spec which improves performance over time. In contrast, the one-shot approach relies on a single prompt that encodes explicit detail about DRAMPyML’s modeling conventions, since the model cannot discover these through iterative inspection of its environment.

\vspace{-0.3cm}
\section{Experiments}\label{sec:experiments}\vspace{-0.2cm}
We now discuss the validation procedure for our proposed Autoformalization agent. We will first discuss our experimental setup, and then continue with an in-depth discussion of our results.
\vspace{-0.2cm}
\subsection{Setup}\vspace{-0.2cm}
\textbf{Dataset}\label{sec:dataset}
We manually constructed DRAMPyML formalizations for 13 JEDEC DRAM standards spanning HBM2-3, LPDDR2-5, DDR2-5, and GDDR5-7. The resulting ground-truth evaluation dataset will be made available on GitHub. Producing such formalizations is labor-intensive, often requiring hours to days per specification due to the need for careful interpretation and modeling of structural and timing constraints. The official JEDEC specification documents vary in length and complexity; DDR2 (2003) has $128$ pages, DDR5 (2020) has $496$ pages, a similar trend can be observed across other DRAM families. Notably, due to the length of JEDEC specifications, full one-shot prompting is infeasible for models with context windows below $10^6$ tokens; however, Sonnet 4.5 supports including the complete specification (excluding figures) in context.

 \textbf{Evaluation metrics}
  \label{sec:evaluation}
As no datasets or standard evaluation metrics exist for hardware autoformalization agents, we introduce two metrics: the Jaccard index on bounded traces and the Timing Constraint Recall score.
 
\textit{Jaccard Index:}
All of the protocol information in a DRAMPyML model $\mathcal{N}$ is contained in the legal command traces $\operatorname{Tr}_k(\mathcal{N})$ that it generates (where $k$ indicates trace length). We compute the Jaccard index on command sequences of depth $k$ to measure similarities between generated Petri nets $\mathcal{N}_{\mathrm{gen}}$ and ground truth $\mathcal{N}_{\mathrm{gt}}$:
\begin{equation}
\label{eq:jaccard-index}
\operatorname{Jacc}_k(\mathcal{N}_{\mathrm{gen}}, \mathcal{N}_{\mathrm{gt}}) = \frac{|\operatorname{Tr}_k(\mathcal{N}_{\mathrm{gen}}) \cap \operatorname{Tr}_k(\mathcal{N}_{\mathrm{gt}})|}{|\operatorname{Tr}_k(\mathcal{N}_{\mathrm{gen}}) \cup \operatorname{Tr}_k(\mathcal{N}_{\mathrm{gt}})|}.
\end{equation}
Inspired by Conjecture \ref{thm:min-config}, we choose $k=4$ and set all device configurations to the minimal setting ($B=2$, $R=1$) during evaluation.

\textit{Timing Constraint Recall:} Second, beyond structural equivalence of un-timed command traces, DRAM specifications also encode \emph{timing constraints} between commands. We define $\operatorname{TC}(\operatorname{PN})$ as the set of timing constraints extracted from a Petri net, where each constraint is characterized by a command pair $(c_{\mathrm{src}}, c_{\mathrm{dst}})$ and a timing expression $\tau$. The timing constraint recall measures the fraction of ground truth constraints recovered by the generated model:
\begin{equation}\label{eq:timing-recall}
\operatorname{TCRecall}(\mathcal{N}_{\mathrm{gen}}, \mathcal{N}_{\mathrm{gt}}) = \frac{|\operatorname{TC}(\mathcal{N}_{\mathrm{gen}}) \cap \operatorname{TC}(\mathcal{N}_{\mathrm{gt}})|}{|\operatorname{TC}(\mathcal{N}_{\mathrm{gt}})|}.
\end{equation}
This requires exact matches on both the command pair and timing expression. Again we use the recall here, because there are timing constraints in the specification documents which are not command-command delays, and so not explicitly included in the manually written Petri net (e.g. write/read latency, preamble timings, and ODT-related timings). Hence we do not penalize the autoformalization agent for including any additional constraints not in ground truth. While one can also evaluate the overlap of the traces of timed command sequences this makes the evaluation more computationally intensive and also provides no insight into the performance of the autoformalization agent at capturing state transition structure and timing related functionality in isolation.

 \begin{figure}[t]
    \vspace{-0.1cm}
    \centering
    \includegraphics[width=\linewidth, trim=0 0 0 2.1cm, clip]{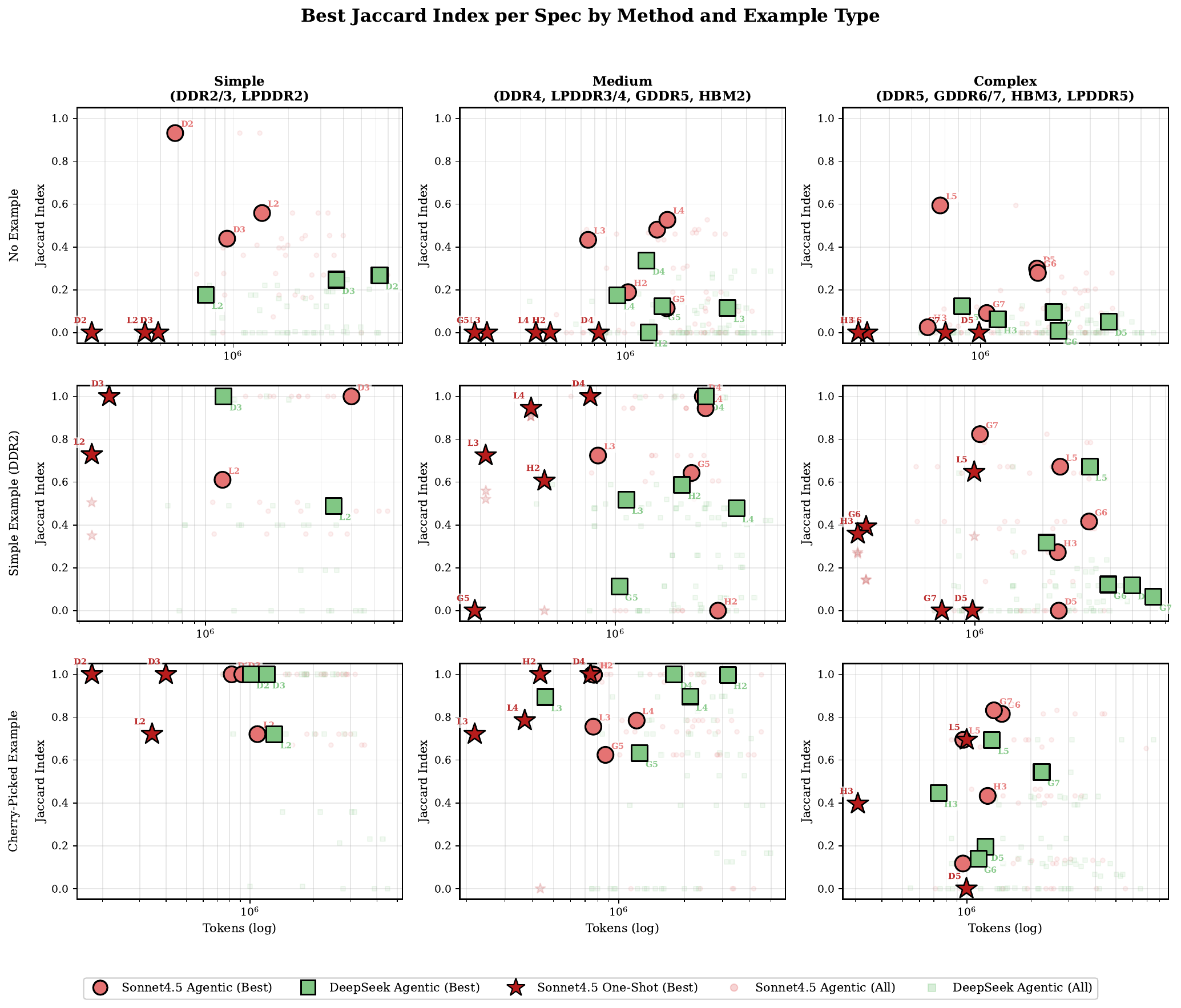}
    \caption{
    % Here we analyse the 
    Pareto-frontier of the Jaccard-index and total token consumption (input+ output) shown for three different priors (no example, simple example and cherry-picked example Petri net) and specs of differing complexity. Bold datapoints show highest performance per specification per family, with the name shortened for brevity (D for DDR, G for GDDR, L for LPDDR and H for HBM). 
    % Agentic generation is generally more performant with weak priors, i.e. no specified example and more complex specifications. One-shotting, is usually more token-efficient but performs comparatively poorly with weaker priors and complex specs as can be seen in the first row and last column. 
    }
    \label{fig:pareto_frontier}
    \vspace{-0.6cm}
\end{figure}

\textbf{Considered Models} We evaluated three models in our agentic autoformalization pipeline: two closed-source (Sonnet 4.5, GPT-5.2) and one open-weight model (DeepSeek-v3p2 \cite{deepseekai2025deepseekv32pushingfrontieropen}). Performance was measured against 13 ground-truth Petri nets using structural $k=4$ Jaccard overlap of untimed command sequences (Eq. \ref{eq:jaccard-index}) and timing constraint recall (Eq. \ref{eq:timing-recall}). Each specification was tested under three prior conditions: no example, a simple example (DDR2), or a cherry-picked example. Cherry-picked examples are adjacent generations within the same family---ideally the previous generation (e.g., DDR3 for DDR4)---mirroring a realistic verification workflow where prior-generation specifications are often available.
\vspace{-0.2cm}
\subsection{Results}\label{sec:results}\vspace{-0.2cm}
We first visualize the Pareto frontier between Jaccard similarity and total token consumption (input+output) for the best-performing open- and closed-source models in Fig. \ref{fig:pareto_frontier}. Without examples, the agentic approach exhibits emergent capability: it produces near-correct Petri nets for DDR2 despite never being exposed to correct examples of our custom DRAMpyML language, and achieves reasonable structural similarity ($>0.5$ Jaccard) even for complex standards such as LPDDR5. In contrast, one-shot generation struggles with the unfamiliar output format under weak priors. In this no-example regime, we also observe a weak scaling trend where performance increases with token usage, a pattern that largely disappears once stronger priors are provided.

\textbf{Providing examples} Providing examples substantially improves results across all approaches. With simple or cherry-picked examples, one-shot generation becomes competitive and can occasionally outperform the agentic method (e.g., HBM2/3 with Sonnet 4.5). Overall, the agentic approach achieves the highest peak performance, particularly for weak priors and more complex specifications, while one-shot generation remains consistently more token-efficient for shorter specs (HBM2/3, LPDDR4, DDR4). With example Petri nets, the agent can construct structurally perfect nets for HBM2 and DDR2–4, though it still struggles with more complex modern specifications with features such as self-refresh in DDR5. It is noteworthy to observe that in this setting the agent can generates Petri nets that deteriorate in quality with increasing token usage as hallucinated tests and verification procedures degrade overlap with the intent of the specification document.

\begin{figure}
    \vspace{-0.2cm}
    \centering
    \includegraphics[width=\linewidth, trim=0 0 0 1.5cm, clip]{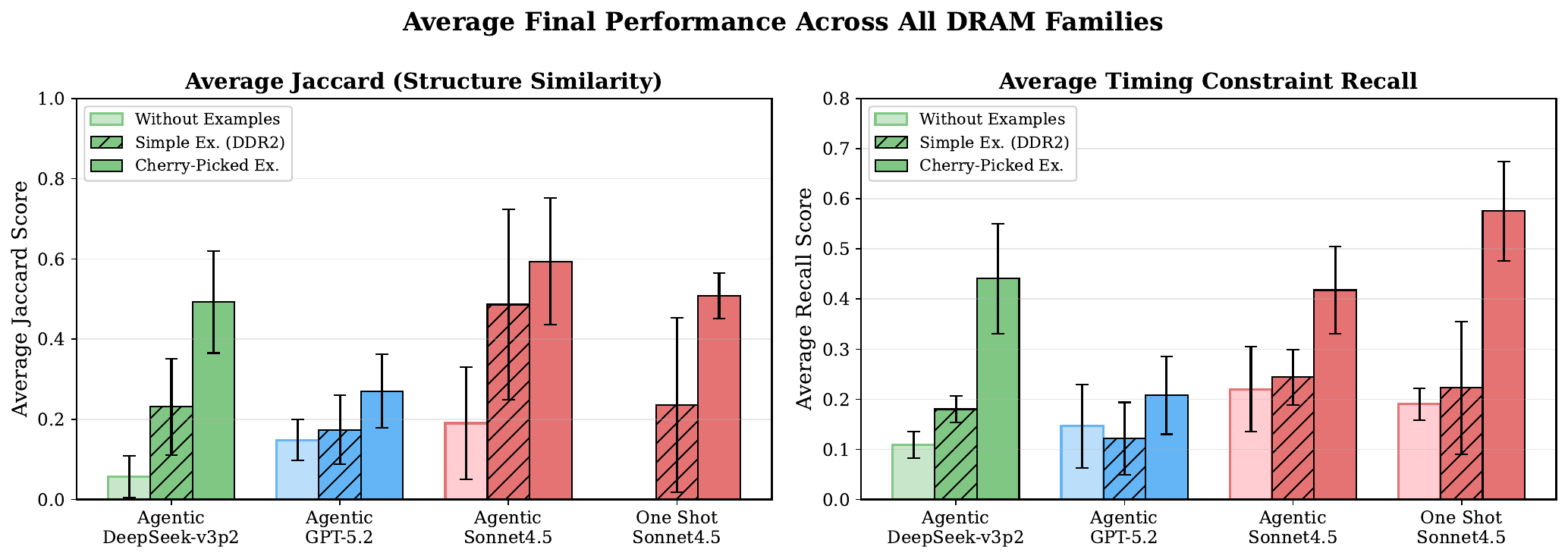}
    \caption{Average Jaccard index and timing-constraint recall relative to ground truth showing that stronger priors—provided via simple or curated example Petri nets—consistently improve performance. DeepSeek-V3.2 \cite{deepseekai2025deepseekv32pushingfrontieropen} is competitive with closed-source models. }
    \label{fig:average_final_performance}
    \vspace{-0.5cm}
\end{figure}

\textbf{Relative Model performance} DeepSeek performs strongly in the agentic setting, surpassing GPT-5.2 on both Jaccard similarity and timing recall, and remaining competitive with the best closed-source model. Due to limited context windows, we could not benchmark DeepSeek or GPT-5.2 on one-shot generation across all specifications. All configurations were run three times, as agentic runs show substantial inter-run variance. Averaged results across specifications are summarized in Fig. \ref{fig:average_final_performance}. More detailed results per specification are shown in App. \ref{sec:additional-results}.
\vspace{-0.2cm}
\subsubsection{Failure Analysis for Complex Specs}
\vspace{-0.2cm}
As a representative failure case on a complex specification, we analyze an LLM-generated DDR5 Petri net with low similarity to ground truth ($k=4$, Jaccard $\approx 0.20$, timing recall $\approx 0.15$). Although the errors are subtle, they severely degrade the correctness of the command sequences. The agent selects the wrong precharge primitive (\texttt{PRESB} instead of per-bank \texttt{PREpb}), omits all \texttt{BL32} command transitions (\texttt{RD32}, \texttt{WR32}, \texttt{RDA32}, \texttt{WRA32}), and fails to include the $\mathrm{PWR}_{\mathrm{ON}}$ state place required for correct conditional execution of \texttt{PDE}/\texttt{PDX}. It also implements an unnecessarily complex dual FAW pool architecture rather than the single global pool. Timing behavior is further compromised by incorrect coordinate query functions: rank-, bank-, and other query functions have internal signatures that do not align with the Petri net coordinates, causing most timing constraints to become inactive and allowing violations to pass undetected. The ground truth avoids this failure mode through a flatter coordinate structure consistent consistent with the Petri net place coordinate structure. Some timing constraints are also completely omitted.

Overall, this example highlights that current agentic systems struggle not only with missing commands/places but also with modeling memory hierarchy and enforcing timing at the correct abstraction level. Future work might consider incorporating stronger structural priors, and memory-hierarchy aware validation tests that generate more accurate models from complex specifications.

\section{Conclusion}
\vspace{-0.2cm}
In this work, we introduced an open-source benchmark for hardware autoformalization covering 13 JEDEC memory standards across DDR, LPDDR, HBM, and GDDR families encouraging further work that pushes the state of the art. We examined theoretical properties of DRAMPyML models and developed efficient procedures for verifying symmetry and correctness of generated formalizations. We further presented an agentic autoformalization framework that outperforms baseline one-shot methods, achieving perfect structural models for several specifications and characterizing the token–performance tradeoff across frontier models.

Forthcoming work will evaluate DRAMPyML more explicitly as an intermediate representation for generating downstream verification collateral such as SVAs, enabling direct comparison with assertion-centric approaches (e.g., \cite{assertllm}). We also plan to incorporate an automated parameter extraction stage into the autoformalization pipeline (Fig.~\ref{fig:autoform_pipeline}), as well as exploring richer languages and further methods for formalizing a wider variety of hardware specifications.


\bibliography{iclr2026_conference}

\begin{thebibliography}{18}
\providecommand{\natexlab}[1]{#1}
\providecommand{\url}[1]{\texttt{#1}}
\expandafter\ifx\csname urlstyle\endcsname\relax
  \providecommand{\doi}[1]{doi: #1}\else
  \providecommand{\doi}{doi: \begingroup \urlstyle{rm}\Url}\fi

\bibitem[{Bai} et~al.(2025){Bai}, {Bany Hamad}, {Suhaib}, and {Ren}]{assertionforge}
Yunsheng {Bai}, Ghaith {Bany Hamad}, Syed {Suhaib}, and Haoxing {Ren}.
\newblock {AssertionForge: Enhancing Formal Verification Assertion Generation with Structured Representation of Specifications and RTL}.
\newblock \emph{arXiv e-prints}, art. arXiv:2503.19174, March 2025.
\newblock \doi{10.48550/arXiv.2503.19174}.

\bibitem[Cerny et~al.(2010)Cerny, Dudani, Havlicek, and Korchemny]{assertions-sv}
Eduard Cerny, Surrendra Dudani, John Havlicek, and Dmitry Korchemny.
\newblock \emph{The Power of Assertions in SystemVerilog}.
\newblock Springer, 2010.
\newblock \doi{10.1007/978-1-4419-6600-1}.

\bibitem[Christ et~al.(2025)Christ, Zimmermann, Barbie, Saberi, Yin, and Jung]{drampyml}
Derek Christ, Thomas Zimmermann, Philippe Barbie, Dmitri Saberi, Yao Yin, and Matthias Jung.
\newblock Drampyml: A formal description of dram protocols with timed petri nets.
\newblock In \emph{Design and Verification Conference and Exhibition (DVCON) Europe 2025}, 2025.

\bibitem[DeepSeek-AI(2025)]{deepseekai2025deepseekv32pushingfrontieropen}
DeepSeek-AI.
\newblock Deepseek-v3.2: Pushing the frontier of open large language models, 2025.
\newblock URL \url{https://arxiv.org/abs/2512.02556}.

\bibitem[Fang et~al.(2024)Fang, Li, Li, Yan, Liu, Xie, and Zhang]{assertllm}
Wenji Fang, Mengming Li, Min Li, Zhiyuan Yan, Shang Liu, Zhiyao Xie, and Hongce Zhang.
\newblock Assertllm: Generating and evaluating hardware verification assertions from design specifications via multi-llms, 2024.
\newblock URL \url{https://arxiv.org/abs/2402.00386}.

\bibitem[Foster(2015)]{functional-verification-survey}
Harry~D. Foster.
\newblock Trends in functional verification: A 2014 industry study.
\newblock In \emph{2015 52nd ACM/EDAC/IEEE Design Automation Conference (DAC)}, pp.\  1--6, 2015.
\newblock \doi{10.1145/2744769.2744921}.

\bibitem[Gulati et~al.(2024)Gulati, Ladsaria, Mishra, Sidhu, and Miranda]{gulatietal2024lean4bench}
Aryan Gulati, Devanshu Ladsaria, Shubhra Mishra, Jasdeep Sidhu, and Brando Miranda.
\newblock An evaluation benchmark for autoformalization in lean4.
\newblock In \emph{ICLR Workshop on AI and Formal Methods}, 2024.
\newblock Benchmark of LLM performance on translating informal mathematics into Lean4 formal code.

\bibitem[Hennessy \& Patterson(2019)Hennessy and Patterson]{golden-age-architectures}
John~L. Hennessy and David~A. Patterson.
\newblock A new golden age for computer architecture.
\newblock \emph{Commun. ACM}, 62\penalty0 (2):\penalty0 48–60, January 2019.
\newblock ISSN 0001-0782.
\newblock \doi{10.1145/3282307}.
\newblock URL \url{https://doi.org/10.1145/3282307}.

\bibitem[Jung et~al.(2016)Jung, Rheinl{\"a}nder, Weis, and Wehn]{jung2016}
Matthias Jung, Carl~C. Rheinl{\"a}nder, Christian Weis, and Norbert Wehn.
\newblock Reverse {{Engineering}} of {{DRAMs}}: {{Row Hammer}} with {{Crosshair}}.
\newblock In \emph{Proceedings of the {{Second International Symposium}} on {{Memory Systems}}}, pp.\  471--476, Alexandria VA USA, October 2016. ACM.
\newblock ISBN 978-1-4503-4305-3.
\newblock \doi{10.1145/2989081.2989114}.

\bibitem[Jung et~al.(2017)Jung, Kraft, and Wehn]{junkra_17}
Matthias Jung, Kira Kraft, and Norbert Wehn.
\newblock A {{New State Model}} for {{DRAMs Using Petri Nets}}.
\newblock In \emph{2017 {{International Conference}} on {{Embedded Computer Systems}}: {{Architectures}}, {{Modeling}}, and {{Simulation}} ({{SAMOS}})}, pp.\  221--226, Pythagorion, Greece, July 2017. IEEE.
\newblock ISBN 978-1-5386-3437-0.
\newblock \doi{10.1109/SAMOS.2017.8344631}.

\bibitem[Jung et~al.(2019)Jung, Kraft, Soliman, Sudarshan, Weis, and Wehn]{junkra_19}
Matthias Jung, Kira Kraft, Taha Soliman, Chirag Sudarshan, Christian Weis, and Norbert Wehn.
\newblock Fast {{Validation}} of {{DRAM Protocols}} with {{Timed Petri Nets}}.
\newblock In \emph{Proceedings of the {{International Symposium}} on {{Memory Systems}}}, pp.\  133--147, Washington District of Columbia USA, September 2019. ACM.
\newblock ISBN 978-1-4503-7206-0.
\newblock \doi{10.1145/3357526.3357556}.

\bibitem[Kim et~al.(2014)Kim, Daly, Kim, Fallin, Lee, Lee, Wilkerson, Lai, and Mutlu]{rowhammer}
Yoongu Kim, Ross Daly, Jeremie Kim, Chris Fallin, Ji~Hye Lee, Donghyuk Lee, Chris Wilkerson, Konrad Lai, and Onur Mutlu.
\newblock Flipping bits in memory without accessing them: an experimental study of dram disturbance errors.
\newblock In \emph{Proceeding of the 41st Annual International Symposium on Computer Architecuture}, ISCA '14, pp.\  361–372. IEEE Press, 2014.
\newblock ISBN 9781479943944.

\bibitem[Shih et~al.(2025)Shih, Lin, Gupta, and Malik]{shih2025flagformalllmassistedsva}
Yu-An Shih, Annie Lin, Aarti Gupta, and Sharad Malik.
\newblock Flag: Formal and llm-assisted sva generation for formal specifications of on-chip communication protocols, 2025.
\newblock URL \url{https://arxiv.org/abs/2504.17226}.

\bibitem[Spear \& Tumbush(2012)Spear and Tumbush]{spear_systemverilog}
Chris Spear and Greg Tumbush.
\newblock \emph{SystemVerilog for Verification}.
\newblock Springer, 2012.

\bibitem[Steiner et~al.(2022{\natexlab{a}})Steiner, Sudarshan, Jung, Stoffel, and Wehn]{dramml}
Lukas Steiner, Chirag Sudarshan, Matthias Jung, Dominik Stoffel, and Norbert Wehn.
\newblock A framework for formal verification of dram controllers.
\newblock In \emph{Proceedings of the 2022 International Symposium on Memory Systems}, MEMSYS 2022, pp.\  1–7. ACM, October 2022{\natexlab{a}}.
\newblock \doi{10.1145/3565053.3565059}.
\newblock URL \url{http://dx.doi.org/10.1145/3565053.3565059}.

\bibitem[Steiner et~al.(2022{\natexlab{b}})Steiner, Sudarshan, Jung, Stoffel, and Wehn]{dramml-svas}
Lukas Steiner, Chirag Sudarshan, Matthias Jung, Dominik Stoffel, and Norbert Wehn.
\newblock A {{Framework}} for {{Formal Verification}} of {{DRAM Controllers}}.
\newblock In \emph{Proceedings of the 2022 {{International Symposium}} on {{Memory Systems}}}, pp.\  1--7, Washington DC USA, October 2022{\natexlab{b}}. ACM.
\newblock ISBN 978-1-4503-9800-8.
\newblock \doi{10.1145/3565053.3565059}.

\bibitem[{Tian} et~al.(2025){Tian}, {Ci}, {Yang}, {Li}, and {Lyu}]{assertcoder}
Enyuan {Tian}, Yiwei {Ci}, Qiusong {Yang}, Yufeng {Li}, and Zhichao {Lyu}.
\newblock {AssertCoder: LLM-Based Assertion Generation via Multimodal Specification Extraction}.
\newblock \emph{arXiv e-prints}, art. arXiv:2507.10338, July 2025.
\newblock \doi{10.48550/arXiv.2507.10338}.

\bibitem[{Wu} et~al.(2025){Wu}, {Li}, {Hu}, {Lin}, {Zhao}, {Wang}, and {Guo}]{shift-left}
Xinyue {Wu}, Zixuan {Li}, Fan {Hu}, Ting {Lin}, Xiaotian {Zhao}, Runxi {Wang}, and Xinfei {Guo}.
\newblock {Shift-Left Techniques in Electronic Design Automation: A Survey}.
\newblock \emph{arXiv e-prints}, art. arXiv:2509.14551, September 2025.
\newblock \doi{10.48550/arXiv.2509.14551}.

\end{thebibliography}
\bibliographystyle{iclr2026_conference}
\clearpage
\appendix
\section{Appendix}
\subsection{Background on DRAM architecture}

\begin{figure}
    \centering
    \includegraphics[width=0.9\linewidth]{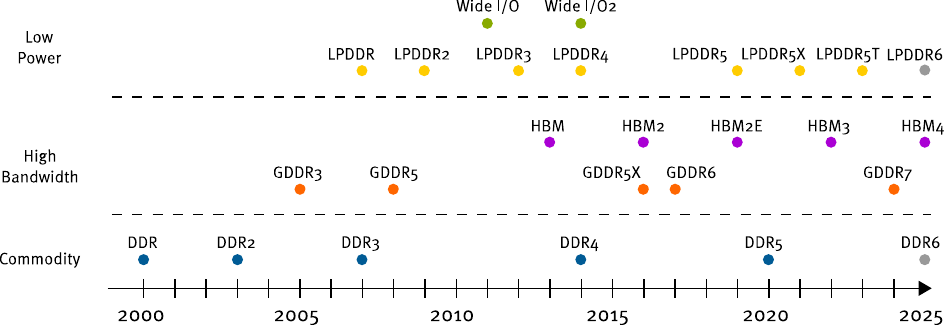}
    \caption{JEDEC standard releases over time, adopted from  \cite{junkra_19}.}
    \label{fig:jedec_standards}
\end{figure}

\begin{figure}[!b]
    \center
    \includegraphics[width=0.8\linewidth]{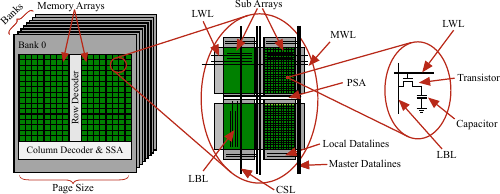}
    \caption{DRAM architecture, showing the organization from a single memory cell up to memory banks~(\cite{jung2016}).}
    \label{fig:dram}
\end{figure}

A DRAM device is a random-access memory (RAM) technology in which each memory cell stores bit information as charge in a capacitor, accessed through an associated transistor. As stored charge leaks away over time, DRAM is dynamic and requires periodic refresh operations to preserve data integrity.

These cells are connected to their local wordline (LWL), which selects an entire row of cells, and local bitline (LBL), which carries the small charge signal from a cell to the sensing circuitry. Cells are densely organized into two-dimensional arrays of rows and columns, and multiple such arrays form a complete memory bank.

Accessing DRAM is fundamentally destructive: reading a cell partially discharges its capacitor, so the sensed value must be restored immediately. This is handled by sense amplifiers, which detect the minute voltage difference on a bitline and drive it to a full digital level. Bitlines are long and highly capacitive, thus sense amplification is one of the dominant contributors to DRAM latency and energy cost.

In each bank, only one row can be activated at a time. When a row is opened, its contents are fetched into the so-called primary sense amplifiers (PSAs), effectively forming a row buffer. Subsequent accesses to the same open row (row hits) are significantly faster than accesses requiring a different row activation (row conflicts), which incur additional precharge and activation delays.

From the PSAs, secondary sense amplifiers (SSAs) and peripheral I/O logic burst data over the shared data bus using a fixed burst length, enabling high bandwidth transfers. Modern DRAM devices further divide banks into multiple bank groups to increase parallelism, though constraints on shared internal resources still limit the number of truly simultaneous operations.

The general architecture of a DRAM device is shown in Fig.~\ref{fig:dram}.

\clearpage
\subsection{Additional Results}
\label{sec:additional-results}
\begin{figure}
    \centering
    \includegraphics[width=\linewidth]{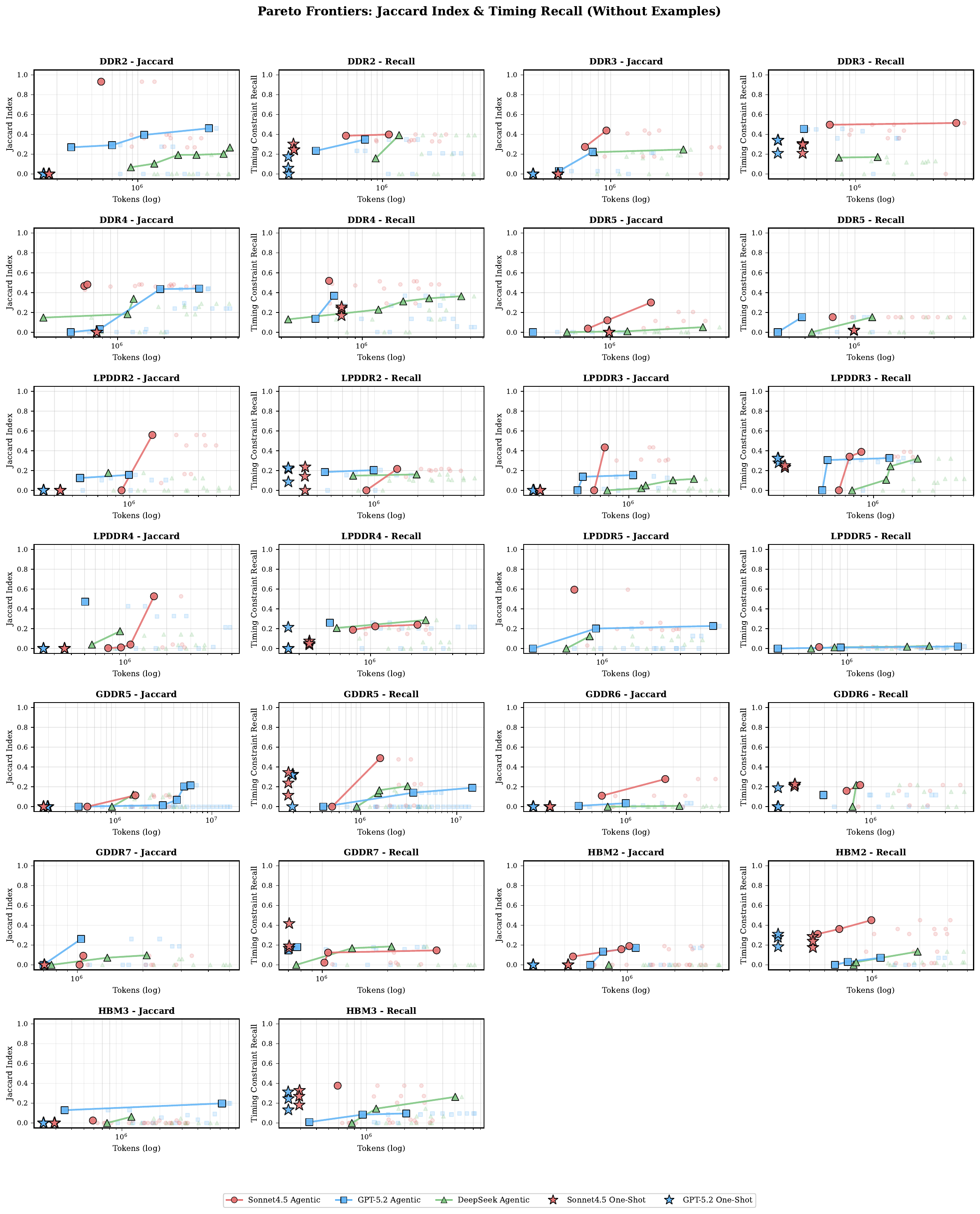}
    \caption{Pareto Frontier of Jaccard Index and Timing Constraint Recall (y) versus tokens (x, log-scale), without examples with respect to ground truth for individual chip specifications.}
    \label{fig:pareto_per_spec_no_example}
\end{figure}

\begin{figure}
    \centering
    \includegraphics[width=\linewidth]{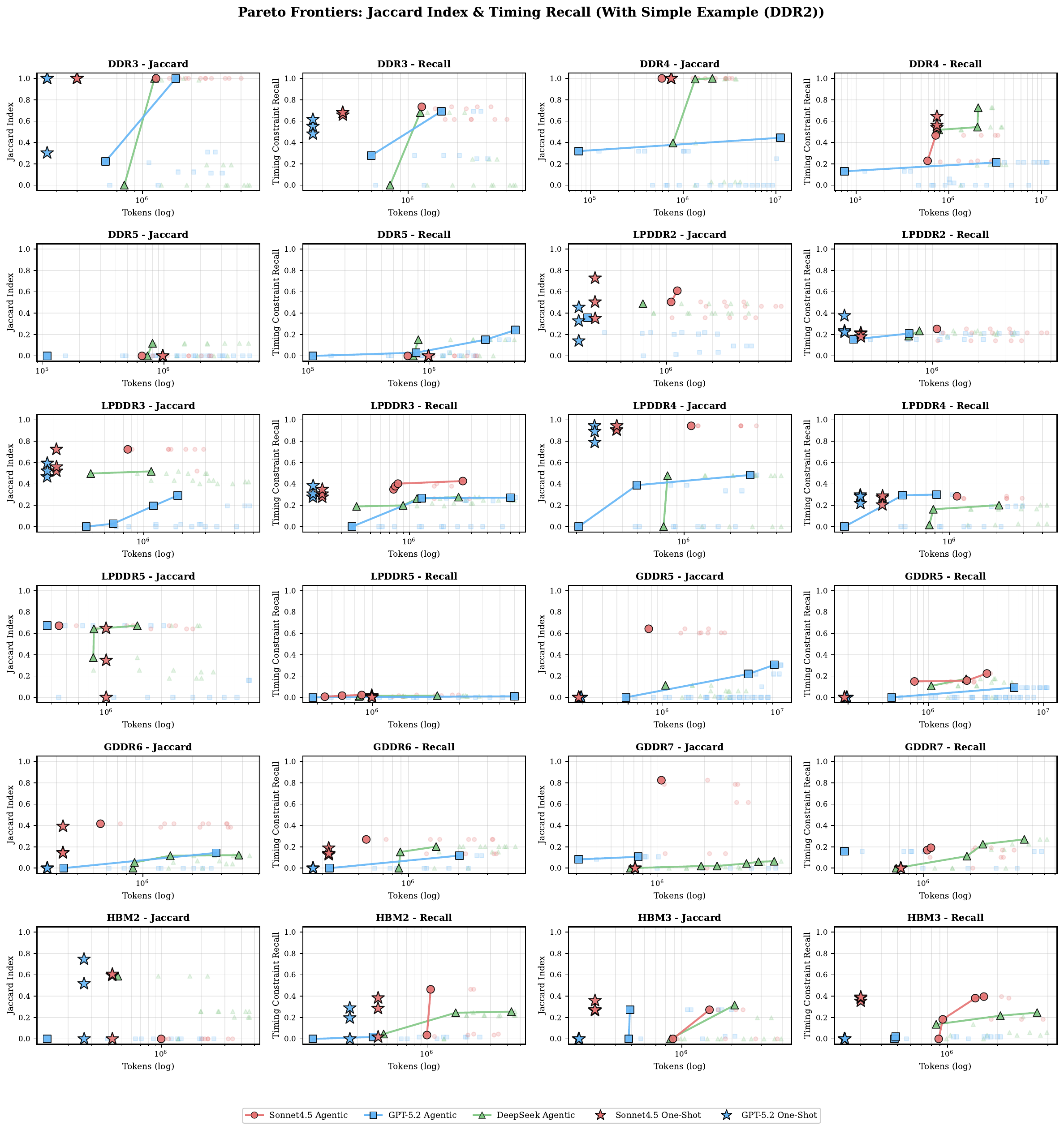}
    \caption{Pareto Frontier of Jaccard Index and Timing Constraint Recall (y) versus tokens (x, log-scale) with respect to ground truth for individual chip specifications with DDR2 example Petri net.}
    \label{fig:pareto_per_spec_simple_example}
\end{figure}

\begin{figure}
    \centering
    \includegraphics[width=\linewidth]{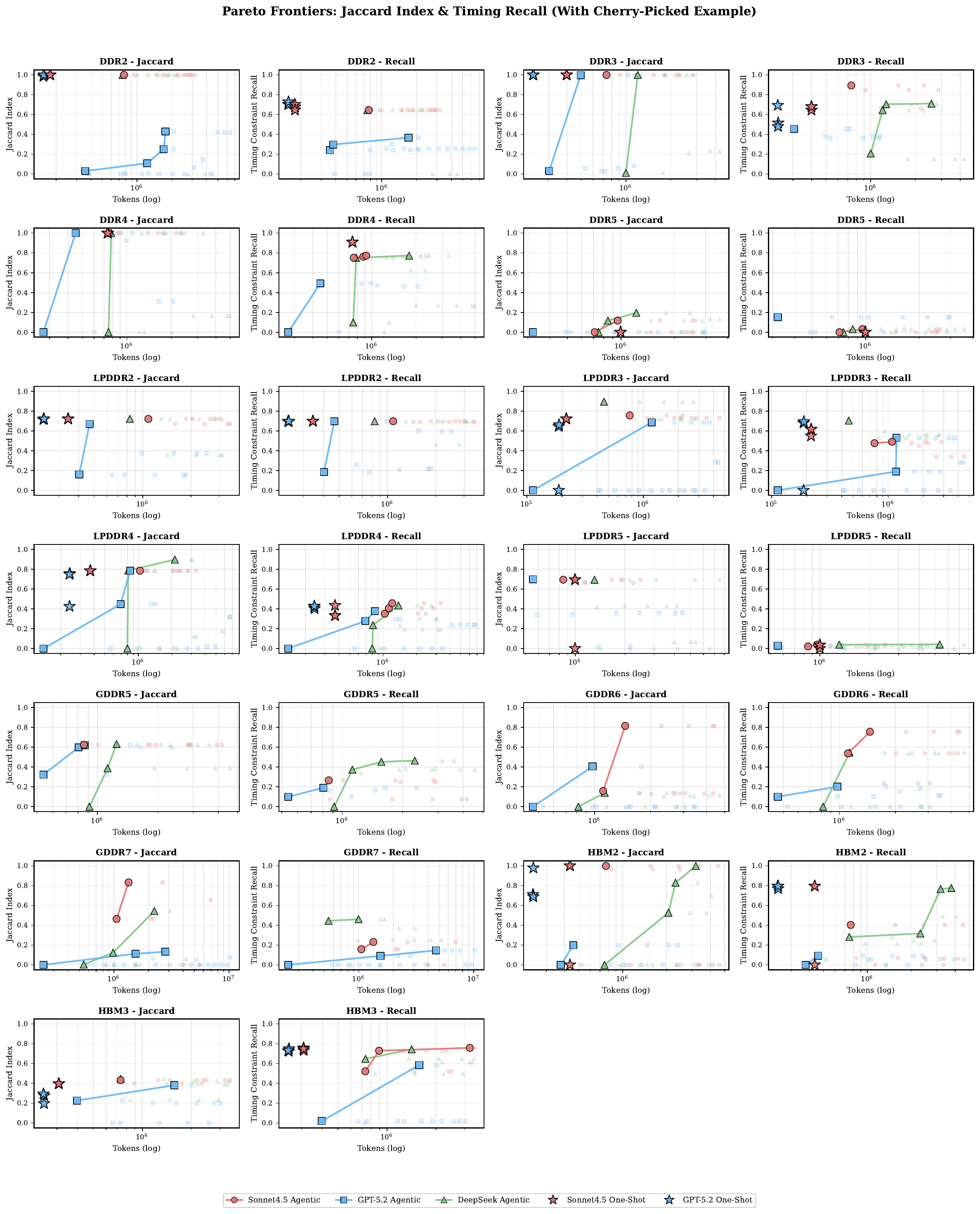}
    \caption{Pareto Frontier of Jaccard Index and Timing Constraint Recall (y) versus tokens (x, log-scale) with respect to ground truth for individual chip specifications with cherry-picked example Petri net.}
    \label{fig:pareto_per_spec_cherry_example}
\end{figure}

All additional results are grouped per spec and presented in Figs. \ref{fig:pareto_per_spec_no_example}, \ref{fig:pareto_per_spec_simple_example} and \ref{fig:pareto_per_spec_cherry_example}.

\end{document}